\documentclass{article}
\pdfoutput=1 

\usepackage[final]{nips_2017}
\usepackage{comment}
\usepackage{booktabs}
\usepackage{amsmath}
\usepackage{graphicx}
\usepackage {titlesec}

\usepackage[utf8]{inputenc} 
\usepackage[T1]{fontenc}    
\usepackage{hyperref}       
\usepackage{url}            
\usepackage{booktabs}       
\usepackage{amsfonts}       
\usepackage{nicefrac}       
\usepackage{microtype}      

\title{Investigating Limit Order Book Characteristics \\ for Short Term Price 
Prediction: \\ a Machine Learning Approach}

\author{
  Faisal Qureshi\\
  Department of Computer Science\\
  University of Toronto\\
  \texttt{fiqureshi1@cs.toronto.edu} \\
}

\begin{document}

\maketitle

\begin{abstract}
With the proliferation of algorithmic high-frequency trading in financial 
markets, the Limit Order Book has generated increased research interest. Research is still at an early stage and there is much we do not understand about the dynamics of Limit Order Books. In this paper, we employ a machine learning approach to investigate Limit Order Book features and their potential to predict short term price movements. This is an initial broad-based investigation that results in some  novel observations about LOB dynamics and identifies several promising directions for further research. Furthermore, we obtain prediction results that are significantly superior to a baseline predictor. 
\end{abstract}

\section{Introduction}
High-frequency trading has rapidly become a major force in the financial markets  and is estimated to account for 55\% of US equity trading volume, 40\% of European equity trading volume, and 80\% of forex futures volume \citep{miller_high_2016}. Changing market dynamics have received considerable commercial and research interest and increased the focus on market micro-structure, high-granularity data and specifically on the Limit Order Book, which is a central element of many high-frequency trading strategies \citep{kirilenko2011flash}, \citep{smith2010high} and \citep{kearns2013machine}.In this study we investigate the various features of the LOB in the specific context of short-term price prediction. Our scope is restricted to predicting the immediate direction of price movement. Prediction of the magnitude of price change, or development of a trading strategy based on such predictions is outside the scope of the current work.  
\paragraph{Limit Order Book}
\label{Limit_Order_Book}
Most modern financial markets facilitate trade through a double auction mechanism centered around a Limit Order Book. Traders submit orders which may be limit orders or market orders. Limit orders specify a price at which they are to be executed. Market orders on the other hand are designed to execute at whatever price is available in the market. The LOB serves as a central record of limit orders waiting to be executed at various price levels. It contains the volume of orders waiting for execution at each price level on both the buy and sell sides. For example, all buy limit orders at the highest price are said to occupy the LOB first level on the buy-side, all buy limit orders at the second highest price are said to occupy the LOB second level on the buy side and so forth. Similarly all sells limit orders at the lowest price are said to occupy the LOB's first level on the sell-side, all sell limit orders at the second lowest price are said to occupy the LOB's second level on the sell side, and so forth. Limit orders at the highest LOB level must be executed before orders at the next LOB level. Within a given LOB level, orders are generally executed on a FIFO basis.\citep{gould2016queue}. From this description it should be evident that the LOB holds much richer multi-dimensional information than price time-series alone, which had been the central focus of price prediction in the past. With the rise of high-speed electronic trading platforms that rapidly disseminate extremely granular (nanosecond) trade information, the LOB plays a central role in modern trading strategies, especially the short time horizon strategies of high-frequency traders. \citep{bonart2017latency}.          

\paragraph{Approach}
We adopt a broad-based approach to study the price predictive potential of the LOB using machine language methods. The master dataset consists of one day of granular (individual order level) data on four large cap NASDAQ stocks (AAPL, AMZN, GOOG, INTC) totaling approximately 800,000 data points. A baseline predictor is devised that makes predictions based on class frequencies - this serves as a baseline for further exploration. The issue of unbalanced classes is encountered  (price remains stationary much more frequently than it moves up or down), and several mitigating approaches are evaluated. Six different classifiers are evaluated on the base dataset and  the best performed is tuned as a benchmark for further study. We are able to achieve predictive results significantly better than baseline. We then systematically investigate the price predictive characteristics of subsets of order and LOB features. In addition to studying the original features of the LOB, we engineer several features based on other successful studies. Our investigation yields some intuitive as well as non-intuitive results.       

\paragraph{Study Outline}
\label{Study Outline}
The rest of this paper is organized as follows: In Section 2 we discuss the source and structure of our data, along with system considerations. In Section 3 we describe the basic optimization of a LOB based price predictor. In section 4 we use our LOB based price predictor to investigate and engineer LOB features. Section 5 outlines our experimental results. In Section 6 we conclude our discussion by highlighting major findings, and identifying areas for future research.      


\section{Data and Systems}
\label{Data_and_Systems}
\paragraph{Data source}
\label{Data_Source}
Major stock exchanges provide market activity data at different levels of granularity. Level 1 data generally consists of best bid/ask prices (a.k.a. best bid/ask quotes) along with last executed price. Level 2 data contains greater detail including information on bid/ask volumes at various price levels. Exchanges also provide very granular data (at the individual order message level) as electronic data feeds format. These data are fairly expensive. Furthermore, the order level data needs to be processed to build a synchronized LOB. Fortunately, the data we required was available from LobsterData (http://Lobsterdata.com) a data provider offering high-quality limit order book data of NASDAQ stocks to the academic community. For this study, we restricted ourselves to using sample data available LobsterData at https://lobsterdata.com/info/DataSamples.php. Specifically, we used the 10-level LOB data for Amazon (AMZN), Apple (AAPL), Google (GOOG), Microsoft (MSFT) and Intel (INTC), each consisting of a single day of trading on June 21, 2012.       
\paragraph{Data Structure}
\label{Data_structure}
As discussed above [\ref{Limit_Order_Book}], the LOB is a listing of the bid/ask volumes available at different price levels. Lobster [\ref{Data_Source}] reconstructs the LOB of NASDAQ stocks from granular order level data. That is, for each order creation, cancellation or execution the LOB is reconstructed. Lobster sample data is available at Lob depths of 1, 5, 10, 30 and 50 respectively. For our purposes, we selected LOB-10 data for the selected stocks. For each stock, the data is organized in a "message file" and "orderbook file". The "message file" contains all order events that affect (in our case) a 10-level LOB. Stated differently, the "message file" contains all order creations, changes, executions and cancellations and deletions that occur at one of the top 10 price levels (on both buy and sell side). The "orderbook file" corresponds row-for-row with the "message file" and provides a reconstructed 10-level Lob for each relevant order event included in the "message file". A schematic of the Lobster file structures is provided in Figure 1 and Figure 2. 

\begin{figure}[htb]
 \label{Figure 1}
  \centering
  \includegraphics[width=0.5\textwidth]{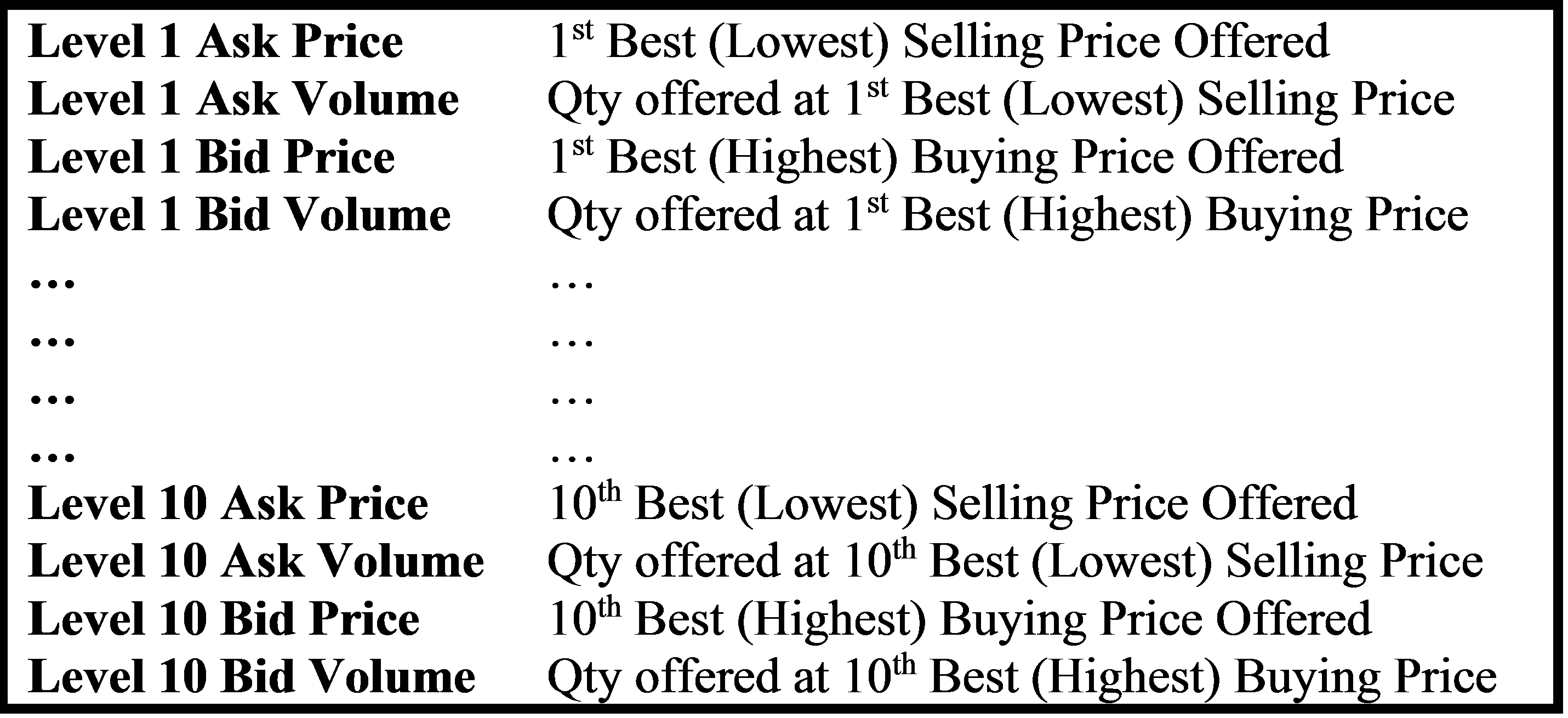}
  \caption{Lobster Orderbook File  Row}
\end{figure}

\begin{figure}[htb]
 \label{Figure 2}
  \centering
  \includegraphics[width=0.5\textwidth]{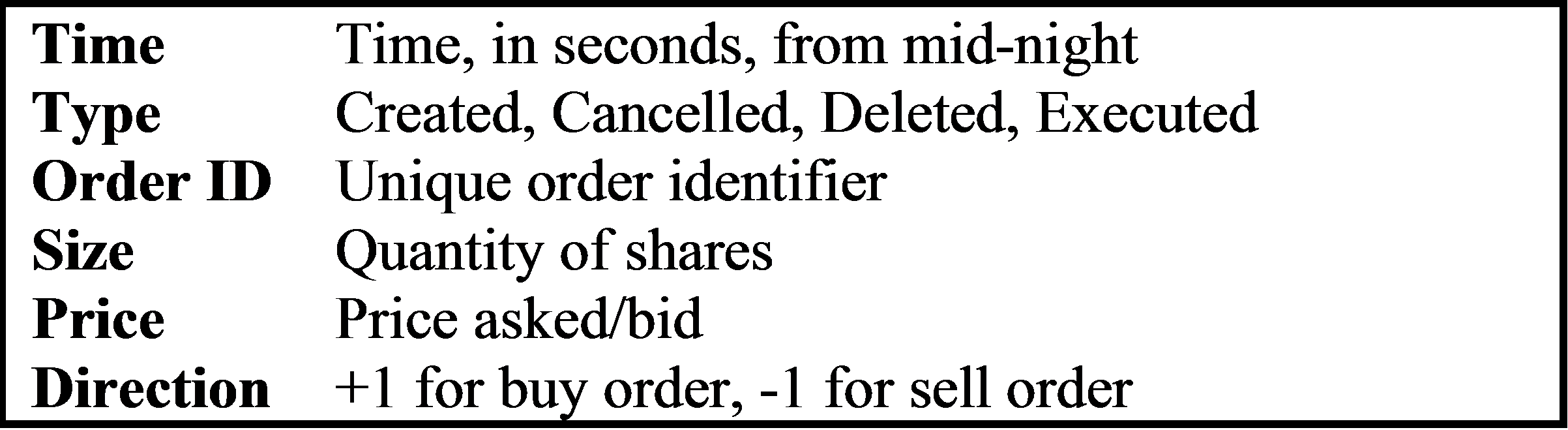}
  \caption{Lobster Message File  Row}
\end{figure}

\paragraph{Computational Considerations}
Study experiments were conducted on an Acer notebook with a 1.83 GHz Intel Celeron processor, 8 Gb RAM operating Windows 8.1 (64 bit) operating system, using Python 3.14 with Spyder IDE. The message and orderbook files for each of the selected stocks range in size from 156,000 to 250,000 lines of data. Due to the volume of data (even for a single day of trading) and computing power limitations, it was not feasible to conduct multiple iterations of the various experiments on the entire dataset. We generally chose to conduct experiments on sample sizes ranging from 5,000 to 50,000 data points. We also conducted the entire series of experiment on the entire dataset for one stock (the program completed in 72 hours). Each time a smaller sample was used to conduct an experiment, we performed a "sanity check" against the results of the corresponding full-data experiment to ensure the results were reasonable. We also performed similar comparisons across the stocks. 

\paragraph{Files, Code, Reports and Logs}
The data files and python scripts used to generate experiments, as well as the experiment logs and results are available online at \url{https://drive.google.com/open?id=1m30OEVc6uZaLOpSeuHSJjpK7X_5qBJBL}.  

\section{Designing a LOB based Price Predictor}
\label{P1: LOB Based Price Predictor}
In the first half of the study, a LOB based price predictor is developed. First price prediction is defined more specifically. Then a baseline predictor "BasePred" is devised which classifies based on label frequency in the test set. BasePred is run against the base dataset to establish performance baseline. Six classifiers are evaluated on the base dataset. A significant class imbalances issue is encountered and we investigate binarization, over/under-sampling and smoothing approaches to counter the issue. The most promising classifier is selected for parameter tuning and further study.
\paragraph{Defining Price Prediction}
In the context of limit orders "sitting in the LOB" at various prices, and order executions occurring at specific prices, the concept of price needs to be defined more specifically. One option is to define price as the last execution price (i.e. last price at which a trade occurred). However, trades occurs much less frequently than order creations, changes and cancellations and much order and LOB activity takes place in between order executions as is evident Figure 2.    Therefore another option is to define price as the mid-point between the best ask and bid price (also known as the "mid-quote price"). We take latter approach in this study. We initially define price prediction as predicting whether (mid-quote) price at the next order event is less than, greater than or equal to the current mid-quote price. Thus at a given time \( t \) (i.e. for a given data point), and mid-quote price at that time \( p_t \), the class label \( C_t \) is defined as follows:
\begin{center}
	\begin{equation}
    	Class \ Label, \, C_t =
    	\begin{cases}
      		-1, & \text{if}\ p_{(t+1)} > p_t \\
      		1, & \text{if}\ p_{(t+1)} < p_t \\
      		0, & \text{otherwise}
    	\end{cases}
	\end{equation}
\end{center}

\paragraph {Metrics}
The main metrics used to evaluate prediction results are accuracy, precision and recall. Since class imbalance issues are anticipated, we shall also calculate class specific accuracy rates. Finally, we shall calculate two versions of the F1 score. F1-weighted provides an F1 score weighted by class frequency and therefore favours the majority class. F1-Micro provides an F1-score as as an unweighted mean of the the class specific F1-scores, treating each class equally. We anticipate a majority C=0 class, and minority C=+1 or -1 classes, but the direction of price change is important, therefore, we shall pay close attention to the F1-Micro score.       

\paragraph{Baseline Predictor}
\label{Baseline_Predictor}
While different classifiers have attained varying levels of predictive success in studies \citep{gould2016queue}, \citep{tsantekidis2017forecasting}, \citep{zheng2012price}, \citep{kercheval2015modelling} and \citep{kearns2013machine}, there is no clear cut winner. Further, the implementations of classifiers and datasets used in these studies were not available to us. We therefore, defined our own basic baseline predictor "BasePred" which predicts class labels randomly with a probability equal to the frequency with which each class appears in the dataset. More specifically, BasePred makes random predictions based on class probabilities \( P(\hat C=k) \), as follows:  
\begin{center}
    \begin{align}
    \begin{split}
    	P(\hat C=k) = \frac {N_{c=k}}{N}
	\end{split}
    \end{align}
\end{center}

where \(N_{c=k} \)  is the number of labels of Class k, \(N\) is the total number of labels, and \(k\) is  in \{ 0,1,-1\}. BasePred achieved high overall accuracy on the Base dataset, however the accuracy for predicting individual classes displayed high variance as described in Table ~\ref{Table 1}. Furthermore, F1-weighted and F1-micro scores varied widely also pointing to a class imbalance issue (discussed in detail below). For comparison purposes a Noise Dataset was generated with random values for all features, and class labels derived from these random values. BasePred performed comparatively on Base and Noise datasets (Table ~\ref{Table 1}) therefore providing a good baseline to ensure evaluated classifiers do not "cheat" by merely modeling class frequencies.  . 

\begin{table}[htb]
  \caption{BasePred Performance on Base and Noise Datasets}
  \label{Table 1}
  \centering
  \begin{tabular}{llllllllll}
    \toprule
    \multicolumn{9}{c}{Stock: AMZN \qquad Classifier: BasePred \qquad Sample Size: 5K}  \\
    \cmidrule{2-9}
    Dataset &Accuracy  &Prec.  &Recall  &F1w  &F1mic  &Acc0 &Acc+1  &Acc-1 \\
    \midrule
    Base  		&78.85	&77.19	&78.85	&78.00	&34.00	&89.51	&7.37	&5.17 \\
    Noise  		&79.39	&79.51	&79.39	&79.45	&33.86	&88.66	&6.74	&6.19 \\
    \bottomrule
  \end{tabular}
\end{table}

\paragraph{Classifier Evaluation}
The following classifiers were selected for evaluation: Guassian Naive Bayes (GNB), Stochastic Gradient Descent (SGD), Multi-layer Perceptron Neural Network (NNet), Support Vector Machine (SVM), Gaussian Process Classifier (GPC) and Random Forest (RF). Standard classifier implementations from the SKlearn library were used. SVM and GPC did not achieve convergence in a reasonable time even for small data samples - therefore these classifiers were eliminated at an early stage. SGD, which models logistic classification or support vector machines under a stochastic gradient descent algorithm, also displayed erratic results. Positive results from RF encouraged us to include ensemble classifiers Adaboost (ADA) and Gradient Boosting Classifier (GDB). Six classifiers were run against the Base dataset, providing the results in Table ~\ref{Table 2}. All classifiers (except SGD) achieved strong overall accuracy, weak class-specific accuracy especially for +1 and -1 classes, and a high variance between F1-weighted and F1-micro scores, reflecting the unbalanced class problem. Furthermore, none performed especially stronger than BasePred, but RF displayed some promise.        

\begin{table}[htb]
  \caption{Six Classifiers on Base Dataset}
  \label{Table 2}
  \centering
  \begin{tabular}{lllllllll}
    \toprule
    \multicolumn{9}{c}{Stock: AMZN \qquad Dataset: Base \qquad Sample Size: 50K}  \\
    \cmidrule{1-9}
    classifier &Accuracy  &Precision  &Recall  &F1w  &F1mic  &Acc0 &Acc+1  &Acc-1 \\
    \midrule
    BasePred	&82.95	&83.04	&82.95	&83.00	&33.67	&44.86	&5.89	&4.28 \\
	GNB		&63.7	&83.84	&63.7	&71.56	&30.34	&33.65	&37.43	&2.01 \\
	SGD		&5.15	&79.17	&5.15	&1.53	&3.35	&0.31	&99.08	&0.00 \\
	NNet	&90.84	&82.52	&90.84	&86.48	&31.73	&49.4	&0.00	&0.00 \\
	ADB		&90.81	&85.52	&90.81	&86.52	&32.08	&49.37	&0.26	&0.27 \\
	RF		&91.21	&89.39	&91.21	&87.75	&39.86	&49.27	&5.10	&8.17 \\
	GDB		&91.05	&88.42	&91.05	&87.23	&36.5	&49.33	&1.44	&6.16 \\
    \bottomrule
  \end{tabular}
\end{table}

\paragraph{Unbalanced Classes}
The initial definition of price prediction (Equation 1) resulted in approximately 90\% of data points belonging to the '0' class, with classes '+1' and '-1' balanced at approximately 5\% each. In other words, from one order event to the next, price remained constant much more frequently than it changed because the majority of order events do not impact the LOB first level. The classifiers (including BasePred) displayed a high overall accuracy and F1-Weighted score but performed not much better than BasePred on F1-Micro and Accuracy of predicting '+1' and '-1'. The Classifiers were modeling the over-frequency of the '0' class. To further investigate the issue, a Binarized dataset was generated by removing all '0' class data-points. The six classifiers were run against the more balanced Binarized dataset resulting in consistent accuracy across classes, reduction in gap between F1-Weighted and F1-Micro , and overall superior performance to BasePred across all classifiers (~\ref{Table 3}). The ensemble classifiers displayed the strongest performance, with RF in a slight lead.  

The Binarized Dataset captures the direction of price movement when price changes but ignores all points in time where price remains stationary, and as such is not applicable to our problem. The synthetic minority over-sampling technique ("SMOTE") is an academically accepted method to address class imbalance issues \citep{chawla2002smote}. The SKlearn Imblearn implementation of SMOTE was used to create an over-sampled (on minority class) and under-sampled (on majority class) version of the Base dataset called the SMOTE dataset. The results of running the six classifiers on the SMOTE dataset are shown in Table ~\ref{Table 4}. SGD and NNet predicted all data as '+1' or '-1' respectively, for reasons not understood and not explored further. GNB continued to have divergent (and not much better than baseline) accuracy across classes. Ensemble classifiers demonstrated stronger overall accuracy, F1-weighted and F1-micro scores compared to BasePred, yet continued to have trouble predicting '+1' and '-1'. The reason for this disparity is discussed in the Smoothing section below.  

\begin{table}[htb]
  \caption{Six Classifiers on Binarized Dataset}
  \label{Table 3}
  \centering
  \begin{tabular}{lllllllll}
    \toprule
    \multicolumn{9}{c}{Stock: AMZN \qquad Dataset: Binarized \qquad Sample Size: 5K}  \\
    \cmidrule{1-9}
    classifier &Accuracy  &Precision  &Recall  &F1w  &F1mic  &Acc0 &Acc+1  &Acc-1 \\
    \midrule
    Rand	&51.12	&51.12	&51.12	&51.12	&51.11	&0.00	&50.13	&52.08 \\
	GNB		&51.72	&54.70	&51.72	&45.07	&45.41	&0.00	&87.33	&17.40 \\
	SGD		&49.07	&24.08	&49.07	&32.31	&32.92	&0.00	&100.0	&0.00 \\
	NNet	&50.93	&25.93	&50.93	&34.37	&33.74	&0.00	&0.00	&100.0 \\
	ADB		&67.00	&67.08	&67.00	&66.99	&66.99	&0.00	&69.14	&64.94 \\
	RF		&71.16	&71.19	&71.16	&71.17	&71.16	&0.00	&71.83	&70.52 \\
	GDB		&70.50	&70.67	&70.5	&70.48	&70.49	&0.00	&73.85	&67.27 \\
    \bottomrule
  \end{tabular}
\end{table}

\begin{table}[htb]
  \caption{Six Classifiers on SMOTE Dataset}
  \label{Table 4}
 \centering
  \begin{tabular}{lllllllll}
    \toprule
    \multicolumn{9}{c}{Stock: AMZN \qquad Dataset: SMOTE \qquad Sample Size: 50K}  \\
    \cmidrule{1-9}
    classifier &Accuracy  &Precision  &Recall  &F1w  &F1mic  &Acc0 &Acc+1  &Acc-1 \\
    \midrule
    BasePred	&33.38	&84.44	&33.38	&45.36	&21.48	&16.42	&32.19	&36.93 \\
	GNB			&61.35	&84.88	&61.35	&70.40	&29.55	&32.18	&2.58	&37.78 \\
	SGD			&4.24	&0.18	&4.24	&0.34	&2.71	&0.00	&100.0	&0.00 \\
	NNet		&4.30	&0.54	&4.30	&0.48	&3.74	&0.00	&1.86	&98.86 \\
	ADB			&80.79	&85.59	&80.79	&82.99	&38.21	&42.80	&23.61	&11.22 \\
	RF			&89.15	&86.22	&89.15	&87.50	&41.34	&47.59	&14.74	&8.10 \\
	GDB			&89.52	&86.43	&89.52	&87.74	&41.72	&47.77	&17.02	&6.53 \\
    \bottomrule
  \end{tabular}
\end{table}

\paragraph{Smoothing}
Over/under-sampling improved results for all classifiers, but did not address the root cause of the unbalanced classes due to defining price prediction in terms of price at next time-step. Price moves in a discontinuous and abrupt manner at each time-step, and even if there is a short term trend in a specific direction, individual arriving orders may move price in either direction, or not move price at all. The classifiers are better at predicting whether price changes or not (i.e. '0' class accuracy) because this is the over-whelming majority case, but are not as good at predicting price change direction (i.e. '+1' and '-1' class accuracy). A way to address the abrupt discontinuous nature of price at each time step is to utilize a smoothing approach similar to \cite{tsantekidis2017forecasting}. First, the mean of the previous \textit{S} prices, denoted by \( \mathit{m_{prev}} \) and the mean of the next \textit{S} prices, denoted by \( \mathit{m_{next}} \) is calculated as follows:     

\begin{center}
    \begin{align}
    	m_{prev} = \frac{1}{S} \sum_{i=o}^{S}p_{t-i} \\
        m_{next} = \frac{1}{S} \sum_{i=o}^{S}p_{t+i}   
    \end{align}
\end{center}

where \( \mathit{p_t} \) is the mid-quote price at time \textit{t}. Then the new smoothed class label \( \mathit{\tilde C_t} \) expresses the direction of price movement at time \textit{t} by comparing the previously calculated quantities \( \mathit{m_{prev}} \) and \(\mathit{m_{next}} \) as follows:

\begin{center}
	\begin{equation}
    	Smoothed \ Class \ Label, \, \tilde C_t =
    	\begin{cases}
      		-1, & \text{if}\ m_b(t) > m_a(t) \cdot (1 + \alpha \ \Delta p_{min}) \\
      		+1, & \text{if}\ m_b(t) < m_a(t) \cdot (1 + \alpha \ \Delta p_{min}) \\
      		0, & \text{otherwise}
    	\end{cases}
	\end{equation}
\end{center}

where \(\Delta p_{min}\) is the day's minimum price change, and \( \alpha \) is a smoothing parameter that defines the least amount price needs to change to be considered as an upward or downward price trend. The six classifiers were run against the new smoothed label dataset assuming \( S=10 \) and \( \alpha = 1\) (Table ~\ref{Table 5}). Smoothing results in a narrowed gap between class-specific accuracy (except in SGD and NNet which respectively fail to predict '0' and '1' completely for reasons not explored further). The ensemble classifiers show markedly superior performance to BasePred, with RF taking a clear lead.  

In the smoothing algorithm the \( \alpha \) parameter defines how sensitive our algorithm is to price changes, and the \textit{S} parameter defines the short-term horizon of backward and forward prices to be considered. These parameters are optimized in the next section after selection of the winning classifier. 

\begin{table}[htb]
  \caption{Six Classifiers on Smoothed Dataset}
  \label{Table 5}
 \centering
  \begin{tabular}{lllllllll}
    \toprule
    \multicolumn{9}{c}{Stock: AMZN \qquad Dataset: Smoothed (S=10, \( \alpha \)=1) \qquad Sample Size: 50K}  \\
    \cmidrule{1-9}
    classifier &Accuracy  &Precision  &Recall  &F1w  &F1mic  &Acc0 &Acc+1  &Acc-1 \\
    \midrule
    BasePred	&34.89	&34.87	&34.89	&34.88	&33.91	&20.88	&28.43	&31.25 \\
	GNB			&38.67	&38.94	&38.67	&38.78	&37.7	&22.55	&31.58	&36.12 \\
	SGD			&27.86	&20.01	&27.86	&12.64	&15.1	&0.00	&99.28	&0.97 \\
	NNet		&38.96	&26.73	&38.96	&30.74	&27.19	&37.68	&0.00	&25.16 \\
	ADB			&51.45	&51.67	&51.45	&51.09	&50.5	&30.69	&42.79	&45.57 \\
	RF			&89.24	&89.24	&89.24	&89.24	&89.41	&43.55	&89.97	&90.63 \\
	GDB			&62.19	&62.85	&62.19	&61.89	&61.4	&36.4	&52.62	&56.13 \\
    \bottomrule
  \end{tabular}
\end{table}

\paragraph{Predictor Selection and Tuning}
RF was a clear winner with the smoothed dataset, and though GDB demonstrated promising results, its runs time was considerably (10X) longer than RF. Therefore, we selected RF as our predictor for the remaining part of the study. We tuned RF using standard K-Fold Validation within the Sklearn GridSearchCV function on the following parameters: number of estimators, maximum features considered and minimum samples per tree leaf. the smoothing algorithms two hyper parameters \( \alpha \) and  \textit{S} were also tuned by running RF on a combined range of values. Results of smoothing parameter tuning are listed in Table ~\ref{Table 6}. Parameters \( \alpha \) = 1 and  \textit{S}=20 were non-rigorously selected as optimal parameters.  This smoothing parametrization produced labels in frequency ratio \( 0:+1:-1 = 47:23:30 \) and produced the best overall and class-specific accuracy. However, prediction of the '0'  class was problematic (<50\%) across the entire parameter range studied. Due to time constraints, we proceeded with the selected parameters for smoothing. Once tuning was complete, we ran RF against the Smoothed dataset Table (~\ref{Table 7}) to establish a benchmark for the next part of the study. 

\begin{table}[htb]
  \caption{Smoothing Parameter Tuning with RF}
  \label{Table 6}
  \centering
  \begin{tabular}{rrrrrrrrrrr}
    \toprule
    \multicolumn{11}{c}{Stock: AMZN \quad Dataset: Smoothed \quad Classifier: RF \quad Sample Size: 50K}  \\
    \cmidrule{2-10}
    S	&\(\alpha\) &Accuracy &F1w  &F1mic  &Acc0 &Acc+1  &Acc-1 &\(N_0\) &\(N_{+1} \) &\(N_{-1}\) \\
    \midrule
    5	&0.5	&74.38	&70.50	&56.53	&46.73	&27.48	&33.65 	&16852	&3750	&4388\\
	5	&1.0	&81.82	&76.91	&48.16	&49.07	&16.54	&17.23 	&19717	&2393	&2880 \\
	5	&1.5	&84.01	&78.69	&42.10	&48.86	&10.50	&9.66 	&20735	&1927	&2328\\
	5	&2.0	&89.23	&84.71	&37.03	&49.42	&5.04	&3.83 	&22138	&1298	&1554\\
	5	&2.5	&91.18	&87.35	&35.81	&49.27	&2.42	&3.88 	&22719	&1047	&1224\\
	10	&0.5	&80.51	&80.27	&78.80	&43.51	&70.75	&72.11 	&12833	&5443	&6704 \\
	10	&1.0	&83.53	&82.54	&75.51	&48.04	&53.16	&64.11 	&16481	&3719	&4780 \\
	10	&1.5	&83.30	&81.37	&68.67	&48.05	&40.48	&49.24 	&18211	&2954	&3815 \\
	10	&2.0	&84.86	&82.21	&62.93	&48.41	&34.74	&33.98 	&19718	&2309	&2953 \\
	10	&2.5	&86.72	&83.67	&57.54	&48.66	&24.36	&28.34 	&20777	&1885	&2318 \\
	20	&0.5	&87.57	&87.48	&87.43	&39.02	&91.17	&91.75 	&8250	&7386	&9324\\
	\textbf{20}	&\textbf{1.0}	&\textbf{87.92}	&\textbf{87.92}	&\textbf{87.81}	&\textbf{42.51}	&\textbf{85.96}	&\textbf{86.91} &\textbf{11720}	&\textbf{5746}	&\textbf{7494} \\
	20	&1.5	&88.89	&88.81	&87.57	&46.44	&83.62	&81.68 &14176	&4683	&6101 \\
	20	&2.0	&89.17	&88.89	&85.93	&47.64	&74.75	&76.97 &16100	&3889	&4971\\
	20	&2.5	&88.76	&88.21	&83.08	&47.15	&65.81	&70.91 &17567	&3254	&4139 \\

    \bottomrule
  \end{tabular}
\end{table}

\begin{table}[htb]
  \caption{RF Performance as Benchmark}
  \label{Table 7}
  \centering
  \begin{tabular}{rrrrrrrrr}
    \toprule
    \multicolumn{9}{c}{Stock: AMZN \quad Dataset: Smoothed (S=20, \( \alpha \)=1) \quad Classifier = RF \quad Sample Size: 100K}  \\
    \cmidrule{1-9}
    Classifier &Accuracy  &Precision  &Recall  &F1w  &F1mic  &Acc0 &Acc+1  &Acc-1 \\
    \midrule
    BasePred	&33.81	&33.82	&33.81	&33.81	&32.90	&19.94	&27.33	&30.83 \\
	RF			&89.47	&89.47	&89.47	&89.47	&89.56	&43.59	&88.91	&91.09 \\
    \bottomrule
  \end{tabular}
\end{table}

\section{Feature Investigation and Engineering}
\label{P2: Feature_Investigation_Engineering}
A LOB-based price predictor was developed in the previous section. In this section, the predictor is used to investigate the price predictive characteristics of various order and LOB features. An approach similar to \cite{kercheval2015modelling} is employed where related features are divided into feature sets to gauge relative prediction informativeness.     

\paragraph{Order Features}
\label{Order Features}
The features in the Lobster message file (Figure 1) were divided into Order Feature Sets as defined in Table ~\ref{Table 8}. The results of running the RF Predictor on Order Feature Sets are listed in Table ~\ref{Table 9}. The Order-Time feature set (consisting of a single feature - time in seconds from mid-night) stands out anomalously with high overall and class-specific accuracy and F1 scores. We believe this may be occurring because we are using a single day of trading data, and the predictor may be over-fitting on the time feature. However, further investigation is required to understand the anomalous behaviour. Surprisingly, Order-ID is a useful feature presumably due to its relationship to Time. Not surprisingly, Order-Direction and Order-Type on their own provide very little information. However, Order-Details (which includes (Size, Price and Direction) is also weakly predictive, perhaps signifying that order information without LOB information is not very informative. 

\begin{figure}[htb]
 \label{Figure 3}
  \centering
  \includegraphics[width=0.5\textwidth]{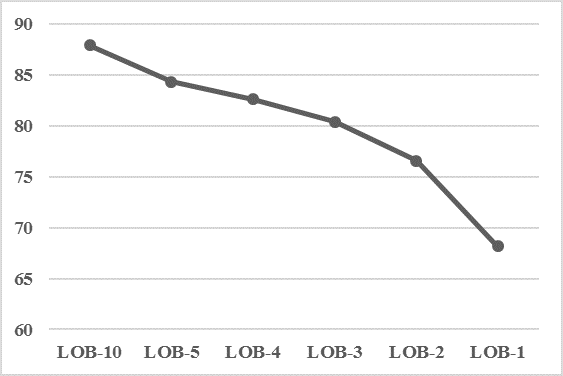}
  \caption{Accuracy vs LOB Levels (Depths)}
\end{figure}

\paragraph{LOB Features}
\label{LOB Features}
The features in the Lobster orderbook file (Figure 1) were divided into LOB Feature sets based on different LOB depths as illustrated in Table ~\ref{Table 10}. The results of running the RF Predictor on LOB Feature Sets are listed in Table ~\ref{Table 11}. As expected, the LOB is strongly predictive. Performance deteriorated at an increasing rate as less LOB levels were used in a Feature Set, with the largest performance drop occurring between LOB-2 and LOB-1, as illustrated in Figure ~\ref{Figure 3}. 

\begin{figure}[htb]
 \label{Figure 4}
  \centering
  \includegraphics[width=0.5\textwidth]{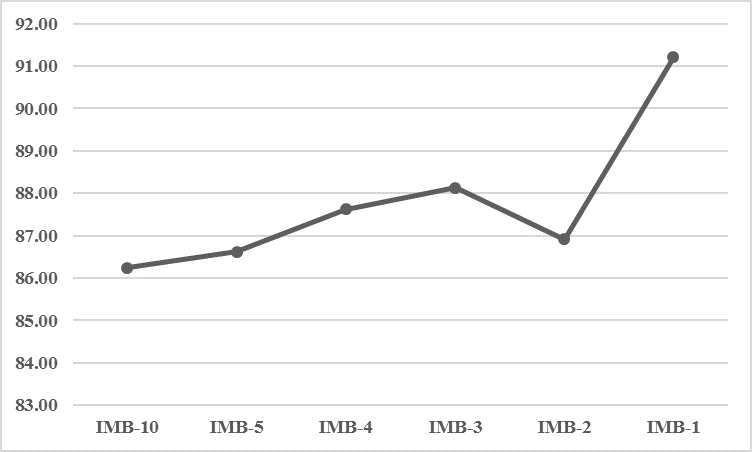}
  \caption{Accuracy vs LOB Imbalance}
\end{figure}

\paragraph{Imbalance Features}
In recent years, several studies \citep{cartea2015enhanced}, \citep{yang2016reduced}, \citep{gould2016queue} have proposed that LOB imbalance is a simple and effective feature for price prediction. The latter study defined LOB imbalance, \( I_{L,t} \), at each LOB price level \(L\) and each time-step \(t\) as follows:

\begin{center}
    \begin{align}
    	I_{L,t} = \frac {V^{buy}_L - V^{sell}_L}{V^{buy}_L + V^{sell}_L}     
    \end{align}
\end{center}

where \(V^{buy}_L\) and \(V^{sell}_L\) are the volume of buy and sell orders, respectively at each LOB level \(L\). Using this definition, LOB imbalance at each of the 10 LOB levels was calculated for the entire dataset, and Imbalance Feature sets were generated as illustrated in Table ~\ref{Table 12}. The results of running the RF Predictor on the Imbalance Feature Sets are listed in Table ~\ref{Table 13}. LOB imbalance features are shown to have predictive value comparable to the LOB itself. Furthermore, while LOB features deteriorate in predictive power as the number of levels is decreased, LOB  imbalance features improve in predictive power as the number of levels is decreased. Finally, the smallest Imbalance Feature Set (IMB-1) is shown to have stronger predictive power than the largest LOB Feature Set (LOB-10). The relationship between Imbalance Level and Accuracy is illustrated in Figure ~\ref{Figure 4}.   

\paragraph{Order Arrival Rates}
The "Order Arrival Rate" was designed as a novel engineered feature, inspired by the success of the order intensity concept in \cite{kercheval2015modelling}. The Order Arrival Rates features capture the volume of buy and sell orders created, canceled and executed within a short historic window \(\Delta t_{hist}\), and also distinguish between orders at LOB level 1 vs all other orders. The Order Arrival Feature Sets are defined in Table ~\ref{Table 14}. The results of running the RF Predictor on Arrival Rate Feature Sets for \(\Delta t_{hist}\) of 0.1, 1.0 and 10.0 seconds are listed in Table ~\ref{Table 15}. Arrival Rates demonstrated some predictive potential. The Arrival Rates of orders at LOB level 1 (Feature Set Arrt-LOBOrds) generally provided better information than the Arrival Rates of other orders (Feature Set Arrt-Ords), except in the peculiar case of \(\Delta t_{hist} = 1.0 sec\). This engineered feature has the potential to outperform non-engineered order features (Table ~\ref{Table 9}) but the \(\Delta t_{hist}\) parameter must be optimized.   

\paragraph{Combining Features}
The best Feature Sets from each category (Orders, LOB, Imbalances, Arrival Rates) were aggregated into a Combined dataset. The results of running the RF Predictor on the Combined dataset are listed in Table ~\ref{Table 16}. However no significant improvement was seen compared to the benchmark of running RF on the Base dataset. Further study is needed to attempt to improve predictive results via feature engineering and feature set optimization.   

\begin{table}[htb]
  \caption{Order Feature Sets}
  \label{Table 8}
  \centering
  \begin{tabular}{ll}
    \toprule
    Feature Set &Features \\
    \midrule
    Orders-All	&Time, Type, ID, Size, Price, Direction \\
	Order-Time	&Time: seconds from mid-night \\
	Order-Type	&Type: create, cancel, execute, delete \\
	Order-Dir	&Direction: +1 for buy, -1 for sell \\
	Order-ID	&ID: unique identifier for each order \\
	Order-Details	&Size, Price, Direction  \\ 
    \bottomrule
  \end{tabular}
\end{table}

\begin{table}[htb]
  \caption{Prediction based on Order Feature Sets}
  \label{Table 9}
  \centering
  \begin{tabular}{lrrrrrrrr}
    \toprule
    \multicolumn{9}{c}{Stock: AMZN \quad Dataset: Smoothed (S=20, \( \alpha \)=1) \quad Classifier = RF \quad Sample: 100K}  \\
    \cmidrule{1-9}
    Feature Set 	&Accuracy  &Precision  &Recall  &F1w  &F1mic  &Acc0 &Acc+1  &Acc-1 \\
    \midrule
    Orders-All		&71.13	&71.12	&71.13	&71.12	&71.22	&34.25	&70.66	&73.97 \\
	Order-Time		&90.83	&90.83	&90.83	&90.79	&90.96	&42.20	&93.63	&93.56 \\
	Order-Type		&39.09	&29.39	&39.09	&23.99	&21.22	&48.46	&0.00	&4.33 \\
	Order-Dir		&38.56	&14.87	&38.56	&21.46	&18.55	&49.77	&0.00	&0.00 \\
	Order-ID		&70.70	&70.74	&70.70	&70.69	&70.66	&34.22	&70.23	&72.98 \\
	Order-Details	&42.77	&42.19	&42.77	&41.26	&40.24	&31.03	&25.29	&34.47 \\
    \bottomrule
  \end{tabular}
\end{table}

\begin{table}[htb]
  \caption{LOB Feature Sets}
  \label{Table 10}
  \centering
  \begin{tabular}{ll}
    \toprule
    Feature Set &Features \\
    \midrule
    LOB-10	&Volume and Price for LOB levels 1 to 10  \\
	LOB-5	&Volume and Price for LOB levels 1 to 5 \\
	LOB-4	&Volume and Price for LOB levels 1 to 4 \\
	LOB-3	&Volume and Price for LOB levels 1 to 3 \\
	LOB-2	&Volume and Price for LOB levels 1 \& 2 \\
	LOB-1	&Volume and Price for LOB level 1 only \\
    \bottomrule
  \end{tabular}
\end{table}

\begin{table}[htb]
  \caption{Prediction based on LOB Feature Sets}
  \label{Table 11}
  \centering
  \begin{tabular}{lrrrrrrrr}
    \toprule
    \multicolumn{9}{c}{Stock: AMZN \quad Dataset: Smoothed (S=20, \( \alpha \)=1) \quad Classifier = RF \quad Sample: 100K}  \\
    \cmidrule{1-9}
    Feature Set 	&Accuracy  &Precision  &Recall  &F1w  &F1mic  &Acc0 &Acc+1  &Acc-1 \\
    \midrule
    LOB-10	&87.92	&87.92	&87.92	&87.91	&88.03	&42.26	&87.98	&90.13 \\
	LOB-5	&84.33	&84.36	&84.33	&84.34	&84.43	&40.68	&83.50	&86.05 \\
	LOB-4	&82.65	&82.66	&82.65	&82.65	&82.71	&40.19	&81.52	&84.35 \\
	LOB-3	&80.43	&80.44	&80.43	&80.43	&80.47	&39.14	&78.94	&82.46 \\
	LOB-2	&76.59	&76.6	&76.59	&76.59	&76.60	&38.11	&75.27	&77.47 \\
	LOB-1	&68.15	&68.16	&68.15	&68.12	&67.97	&34.83	&64.23	&68.32 \\
    \bottomrule
  \end{tabular}
\end{table}

\begin{table}[htb]
  \caption{LOB Imbalance Feature Sets}
  \label{Table 12}
  \centering
  \begin{tabular}{ll}
    \toprule
    Feature Set &Features \\
    \midrule
    IMB-10	&\(I_{L,t}\) for L in [1,10]  \\
	IMB-5	&\(I_{L,t}\) for L in [1,5] \\
	IMB-4	&\(I_{L,t}\) for L in [1,4] \\ 
	IMB-3	&\(I_{L,t}\) for L in [1,3] \\ 
	IMB-2	&\(I_{L,t}\) for L in [1,2] \\ 
	IMB-1	&\(I_{L,t}\) for L = 1 \\ 
    \bottomrule
  \end{tabular}
\end{table}

\begin{table}[htb]
  \caption{Prediction based on LOB Imbalance Feature Sets}
  \label{Table 13}
  \centering
  \begin{tabular}{lrrrrrrrr}
    \toprule
    \multicolumn{9}{c}{Stock: AMZN \quad Dataset: Smoothed (S=20, \( \alpha \)=1) \quad Classifier = RF \quad Sample: 100K}  \\
    \cmidrule{1-9}
    Feature Set 	&Accuracy  &Precision  &Recall  &F1w  &F1mic  &Acc0 &Acc+1  &Acc-1 \\
    \midrule
    IMB-10	&86.24	&86.25	&86.24	&86.24	&86.29	&42.40	&85.59	&87.16 \\
	IMB-5	&86.62	&86.62	&86.62	&86.61	&86.70	&41.42	&86.56	&88.97\\
	IMB-4	&87.62	&87.63	&87.62	&87.61	&87.70	&41.75	&88.44	&89.62\\
	IMB-3	&88.13	&88.13	&88.13	&88.12	&88.23	&41.55	&89.23	&90.10\\
	IMB-2	&86.91	&86.93	&86.91	&86.89	&87.00	&41.25	&88.65	&88.97\\
	IMB-1	&91.21	&91.20	&91.21	&91.18	&91.36	&42.99	&93.45	&93.77\\
    \bottomrule
  \end{tabular}
\end{table}

\begin{table}[htb]
  \caption{Arrival Rate Feature Sets}
  \label{Table 14}
  \centering
  \begin{tabular}{ll}
    \toprule
    Feature Set &Features \\
    \midrule
 ArrRt	&\(V^{buy}_{created},V^{sell}_{created},V^{buy-sell}_{created},\) \\ \\							&\(V^{buy}_{canceled},V^{sell}_{canceled},V^{buy-sell}_{cancelled}, V_{executed} \) \\ \\

ArrRt-Ords &\(V^{buy}_{nonLOB1,created},V^{sell}_{nonLOB1,created},V^{buy-sell}_{nonLOB1,created},\) \\ \\ 	&\(V^{buy}_{nonLOB1,canceled},V^{sell}_{nonLOB1,canceled},V^{buy-sell}_{nonLOB1,cancelled} \) \\ \\

ArrRt-LOBOrds &\(V^{buy}_{LOB1,created},V^{sell}_{LOB1,created},V^{buy-sell}_{LOB1,created},\) \\ \\ 		&\(V^{buy}_{LOB1,canceled},V^{sell}_{LOB1,canceled},V^{bu-ysell}_{LOB1,cancelled} \) \\ \\
    \bottomrule
  \end{tabular}
\end{table}

\begin{table}[htb]
  \caption{Prediction based on Arrival Rate Feature Sets}
  \label{Table 15}
  \centering
  \begin{tabular}{lrrrrrrrrr}
    \toprule
    \multicolumn{10}{c}{Stock: AMZN \quad Dataset: Smoothed (S=20, \( \alpha \)=1) \quad Classifier = RF \quad Sample: 100K}  \\
    \cmidrule{1-10}
    Feature Set &\(\Delta t_{hist}\)(sec.) &Accuracy  &Precision  &Recall  &F1w  &F1mic  &Acc0 &Acc+1  &Acc-1 \\
    \midrule
ArrRt			&0.1	&74.38	&75.28	&74.38	&74.34	&74.26	&40.22	&67.13	&70.90\\
ArrRt-Ords		&0.1	&34.40	&34.37	&34.40	&34.38	&33.50	&20.09	&27.94	&31.56\\
ArrRt-LOBOrds	&0.1	&69.45	&70.46	&69.45	&69.32	&69.09	&38.73	&60.72	&64.72\\
ArrRt			&1.0	&34.31	&34.33	&34.31	&34.32	&33.43	&20.03	&28.36	&31.32\\
ArrRt-Ords		&1.0	&62.85	&65.82	&62.85	&62.33	&61.85	&39.44	&50.07	&51.70\\
ArrRt-LOBOrds	&1.0	&34.02	&33.97	&34.02	&34.00	&33.07	&20.13	&27.78	&30.42\\
ArrRt			&10.0	&84.54	&84.93	&84.54	&84.57	&84.63	&43.43	&80.44	&83.05\\
ArrRt-Ords		&10.0	&34.14	&34.21	&34.14	&34.17	&33.27	&19.94	&28.19	&31.28\\
ArrRt-LOBOrds	&10.0	&80.14	&80.77	&80.14	&80.13	&80.10	&42.64	&73.85	&77.67\\
    \bottomrule
  \end{tabular}
\end{table}

\begin{table}[htb]
  \caption{RF Performance on Base vs Combined Dataset}
  \label{Table 16}
  \centering
  \begin{tabular}{rrrrrrrrr}
    \toprule
    \multicolumn{9}{c}{Stock: AMZN \quad Classifier = RF \quad Sample Size: 100K}  \\
    \cmidrule{1-9}
    Dataset &Accuracy  &Precision  &Recall  &F1w  &F1mic  &Acc0 &Acc+1  &Acc-1 \\
    \midrule
	Base			&89.47	&89.47	&89.47	&89.47	&89.56	&43.59	&88.91	&91.09 \\
    Combined		&89.54	&89.64	&89.54	&89.55	&89.63	&44.24	&87.23	&89.68 \\
    \bottomrule
  \end{tabular}
\end{table}

\clearpage

\section{Conclusion}
In this study, we attempted to develop a LOB based price predictor. Several classifiers were evaluated. Class imbalance in our dataset was addressed relatively effectively via smoothing. Random Forest was determined to be the optimal classifier (from amongst those tested), and after classifier and smoothing parameter tuning, we achieved significantly superior prediction results to a baseline (class frequency based) predictor (Table ~\ref{Table 7}). Our predictor was used to evaluate several feature sets containing various order and LOB features. 

There are several ways in which this study could be improved and several avenues for further exploration. A larger data set including multiple weeks or months and multiple stocks should be incorporated in the analysis. Classifiers which behaved erratically in this study (SGD and NNet) should be looked at further as these have demonstrated positive results in other studies. Classifiers which proved too computationally intensive (GPC, GDB and SVM-nonSGD) should also be looked at to see if tuning can make these classifiers workable. The phenomenon of achieving low accuracy on '0' predictions after smoothing warrants further investigation. The anomalously high prediction accuracy obtainable using the Time feature alone needs to be understood further ~\ref{Order Features}. LOB Features' predictive power deteriorated as less LOB levels were incorporated, however the opposite was true for LOB imbalance features - this phenomenon warrants further investigation. Order Arrival Rates is an engineered feature that demonstrates potential. LOB change rates (the short term historic change in LOB) could be another engineered feature worth studying. Finally, the optimization of LOB feature sets and development of engineered features with strong predictive potential could be avenues of future research.          

\bibliographystyle{plainnat}
\bibliography{References}

\end{document}